# Machine Learning to Predict the *L*-Point Direct Bandgap of Bi$_{1-x}$Sb$_x$ Nanomaterials

Shuang Tang[1,*], Jenna Jean-Baptiste[1], Schuyler Vecchiano[1], Adam Lukasiewicz[1], Alexandria Burger[1]

[1] *College of Engineering, State University of New York Polytechnic Institute, Albany/Utica, NY, 12203/13502, USA*

\* ***E-mail***: *tangs1@sunypoly.edu*

## Abstract

With the development of modern nanoscience and nanotechnology, Bi$_{1-x}$Sb$_x$ can be synthesized into different nanoscale and nanostructured forms, including thin films, nanowires, nanotubes, nanoribbons, and many others. However, due to the strong correlation between electrons and holes at the *L*-point in the Brillouin zone, the direct band evolves in an anomalous manner under the quantum confinement when nanostructured. Due to the alloying and the low symmetry, predicting the *L*-point direct bandgap in a nanomaterial using either *ab initio* calculations or **k·p** perturbations can be computationally costive or inaccurate. We here try to solve this problem using the machine learning methods, including the support vector regression, the regression tree, the Gaussian process regression, and the artificial neural network. A goodness-of-fit of ~0.99 can be achieved for Bi$_{1-x}$Sb$_x$ thin films and nanowires.

## 1. Introduction

The magnitude, degeneracy, and edge-side density of states of the *L*-point direct bandgap in a Bi$_{1-x}$Sb$_x$ nanomaterial can be affected by many factors, including the temperature, the composition, the crystal orientation, and the materials size. This richness of controllability has led to their broad applications in infrared[1, 2], cryogenic [3, 4] and terahertz devices. However, the strong correlation between the electrons and holes at the *L*-point are making it challenging to accurately and efficiently predict the bandgap either with *ab initio* calculations or with **k·p** perturbations [5-8].

With the recent rapid development of artificial intelligence, many black-box and phenomenological tools have been created and accepted by researchers to predict various materials properties in an economical and efficient manner, using machine learning methodologies [9-16], the mechanical strength [17-22]. Owolabi et al. have used support vector (SV) regression to predict bandgaps of doped TiO$_2$ semiconductors [23] and generate the crystal lattice parameters of pseudo-cubic/cubic perovskites [24]. Bagged trees (BT) are used to study band structure of ABX(3) compounds [25], absorption energies of metal alloys[26], fatigue strength of steels [27], and many others [28]. Gaussian process (GP) regressions are used to predict electron-phonon coupling and thermoelectric *ZT* [29], electron inelastic mean free paths [30], and alloy melting temperatures[31]. Besides, bandgaps [32], electron correlations [33], and crystal structures [34] have also been predicted using artificial neural networks (ANN). Recently, Wu et al. employed a comprehensive approach based on the linear Lasso, the ridge regression, the random forest, the SV regression, the GPs and the ANN algorithms to predict the bandgaps of 3896 inorganic semiconductors [35]. However, to our best knowledge, there has not been a machine learning methodology developed yet to predict the $L$-point direct bandgap of $Bi_{1-x}Sb_x$ nanomaterials.

In this paper, we first use the traditional machine learners of SV, BT, and GP regression to explore how to predict the $L$-point direct bandgap of $Bi_{1-x}Sb_x$ thin films and nanowires of different alloying composition, film thickness, wire diameter, and growth orientation. We then implement the ANNs to count the nonlinearity that exists in the mapping from the feature space to the target value. The goodness-of-fit can be as high as ~0.99 when 10 neurons are used in the hidden layer of a two-layer forward ANN. The overfitting issue will also be discussed.

## 2. Methodology

### 2.1 Support Vector Regression[23, 24]

In an SV regression learner, a hyperplane in the space formed by both the feature vector and the target is constructed to conclude the mapping from the feature vector to the target as shown in Figure 1 (a). Unlike in a traditional linear regression, there exists a margin of $\varepsilon$ from the hyperplane, within which all the data points are supposed to be included. The data points that fall right onto the margin edge are called the support vectors (SVs). For a soft margin algorithm, some slack vectors are allowed outside the region of $\varepsilon$. Both the support and the slack vectors are significant in determining the total loss function during the procedure of finding the hyperplane by minimizing:

$$L(\varepsilon, \xi) = \frac{1}{2}\|\boldsymbol{w}\|^2 + \xi \sum_{i=1}^{N} max(0, |y_i - \boldsymbol{w}^T \boldsymbol{x}_i| - \varepsilon),$$

where $\varepsilon$ and $\xi$ are the hyperparameters, **w** is the coefficient vector, $N$ is the size of the training pool, and $\boldsymbol{x_i}$ and $y_i$ are representing the feature vector and the target value of the $i$th data point.

## 2.2 Bagged Tree Regression[25-28]

A regression tree is a greedy algorithm implementing the divide-and-conquer strategy. The data in the training pool are partitioned into separate regions in the feature-target space using a decision tree algorithm, as illustrated in Figure 1 (b), such that the sum of square errors

$$SS = \sum_{C \in leaves(T)} \sum_{i \in C} (y_i - \bar{y}_c)^2$$

is minimized, where $\bar{y}_c$ is the mean of target values inside Region $C$. Within each region, the target can be predicted by the region mean or by a simple regressor that best fits the local samples. The tree architecture is constructed with a recursive splitting rule starting from the root node and ends at the leaf nodes. The recursive splitting continues from a mother node to the child nodes until the minimization is achieved. Unlike the SV regression, the boundary created by a tree between each two partitions is not restricted to be linear, where a piecewise steplike hypersurface is usually obtained to address the approximation of nonlinearity. One drawback of the tree regression is that they are very sensitive to the addition or removing sample data points. Therefore, a bagging procedure is used to reduce such sensitivity. Multiple trees are created and bagged using a controlled bootstrapping to ensure that each tree is deep, intact and of low biases. The overall variance of the prediction is reduced by averaging among different trees.

## 2.3 Gaussian Process Regression[29-31]

Each data point represented as a vector in the feature space can be deterministic or probabilistic. In a natural process, a probabilistic event can be usually described using a Gaussian distribution. Further, in a high-dimensional space, the similarity or difference between two vectors can be described not only by their Euclidean distance, but also by other kernels, such as the angle, the inner product, the radial overlap, or the Gaussian similarity between two vectors. Figure 1 (c) is illustrating how the Gaussian kernel varies with the component differences between two vectors as,

$$K_{ij} = \varsigma e^{-\frac{1}{2}\Sigma_k \left(\frac{x_{i,k}-x_{j,k}}{l_k}\right)^2}$$

where $\varsigma$ is a normalization coefficient, $k$ spans over all the dimensions, and $l_k$ is the diffusion length in the $k$th dimension.

## 2.4 Artificial Neural Network[32-35]

The ANN is a bionic invention inspired by the neuroscience. It mimics the manner of how information is received, processed, and recognized in a biological brain [36-38]. An artificial neuron is a virtual concept that receives real number signals from all inputting units connected to it, combines the signals into a weighted sum, processes the sum using a nonlinear function, and then pass the processed signal to multiple units that are connected to itself in the next layer. During the training procedure, the weights associated with each connection and neuron are updated in a recursive approach, as is mimicking the signal strengthening and weakening phenomena when an animal brain is learning a new pattern. The information processing inside the neurons between the input and the output layers are usually functioning in black-box style, without rendering apparent physical meanings, and are therefore referred to as the hidden layers. One ANN can have one or multiple hidden layers, depending on the complexity of the learning task, as illustrated in Figure 1 (d).

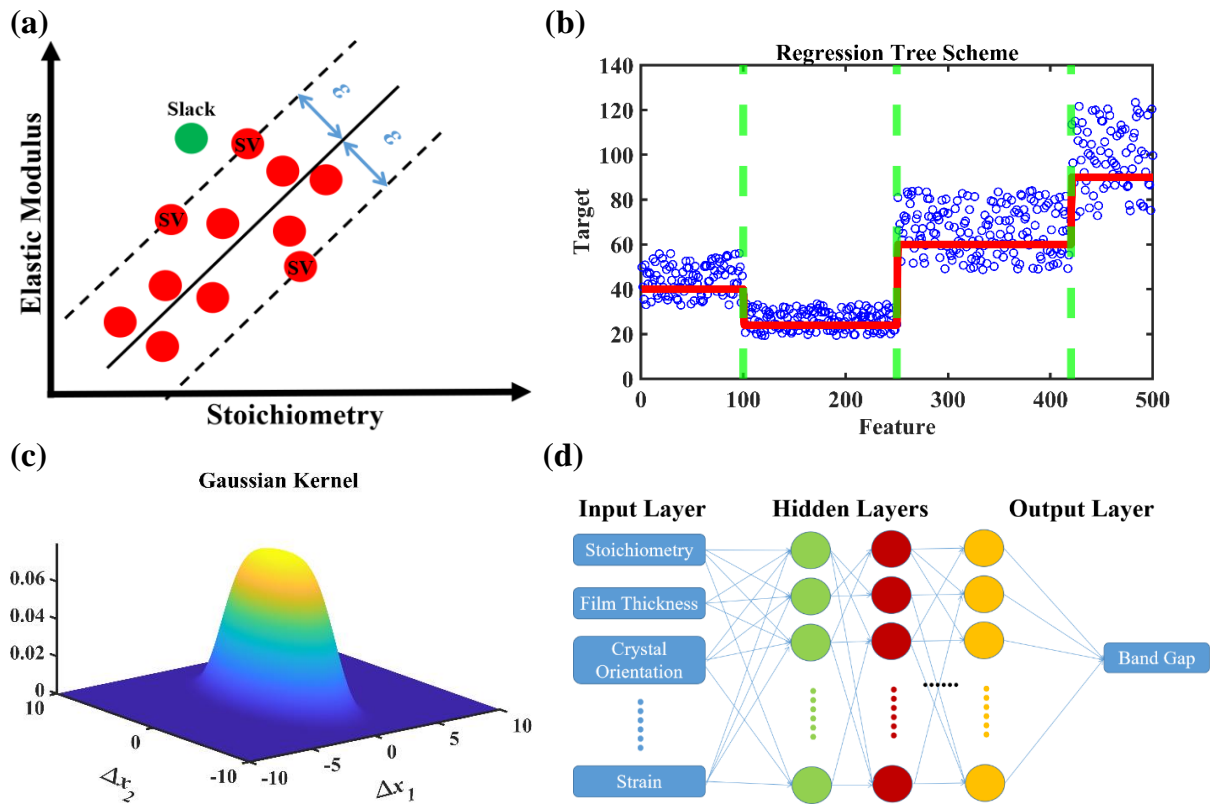

**FIGURE 1:** Illustration of machine learning algorithms for materials property prediction. **(a)** In a support vector regression learner, a hyperplane (solid curve) with a margin of $\varepsilon$ is used to include as many data points into the region as possible. The data points in the feature space on the boundaries of the margins are called the support vectors (SV), and any data points outside are called the slack vectors, both of which are determining the place and shape of the hyperplane. **(b)** A regression tree uses a divide-and-conquer strategy to partition the feature space into separate regions using a decision tree algorithm, such that the total intra-region variance is minimized, as marked by the horizontal red lines and the vertical green lines. The regressor in each region can be then optimized independently and locally. **(c)** In a Gaussian process regression learner, the data points are supposed to have a multidimensional Gaussian distribution. The difference between two data points is usually measured by a Gaussian kernel instead of a Euclidean distance. **(d)** In an artificial neural network learner, the raw information is delivered into the machine through the input layer, and a target value is obtained at the output layer, between both of which, one or multiple hidden layers are used to process the information. The neurons in each layer are connected by weighted and directional edges.

## 3. Results and Discussions

We will first investigate the performance of several commonly used machine learners in predicting the $L$-point bandgap of a $Bi_{1-x}Sb_x$ thin film or nanowire that is grown along the trigonal direction in the rhombohedral crystal lattice. The general growth orientation scenario will be discussed later. The ~100 sample data of our training pool are collected from the previous works [39-42], with different values of alloying composition and film thicknesses or wire diameter as exhibited in Figure 2 (a) and (b). The sample points are randomly distributed within the feature space to ensure that the learning process is not biased. First, we check the performance of SV regression. The prediction accuracy is acceptable when this $L$-point direct bandgap is smaller than ~60 meV for the thin film or ~100 meV for the nanowire, as marked by the green stars in Figure 2 (c) and (d), respectively. However, when the bandgap is further increased at the region of smaller film thickness or thinner wire diameter, the prediction tends to underestimate the target value. As the bandgap increases, this underestimation becomes more significant, which reaches as large as ~80% for a target value of ~400 meV in the thin film predictor and ~75% for a target value of ~2000 meV in the nanowire predictor. This discrepancy at the high bandgap end might be due to the fact that the non-linear correlation between the electrons and holes at the $L$-point can be barely captured by the linear modeling in SV regressions. Further, because the regression relation in an SV regression is mainly determined by the data points at or outside the $\varepsilon$-margin edges, the accuracy for the non-support and non-slack points are sacrificed.

To overcome these drawbacks, we then implement a BT learner to carry out the divide-and-conquer approach, and ensure that the regressions at the small-, the medium, and the large-gap regions do not hinder each other. The prediction results are exhibited by the blue pentagrams in Figure 2 (c) and (d). It seems that the medium-gap section has the best prediction, and both the small- and the large-gap sections suffer from an overestimation. By comparison between the large and the small value ends, the inaccuracy worsens for small bandgaps in the thin film predictor, but is more highlighted for large bandgaps in the nanowire predictor. However, despite this ~50% underestimation at its worst case, the overall accuracy is obviously improved by the tree partitioning. A GP learner is then utilized to address the issues of differentiation and nonlinearity that may exist during the mapping from the feature space to the predicting target, in which the regression around each data point can be carried out with the main emphasize on its relation to adjacent points. As expected, the prediction accuracy is further improved in the whole range of bandgaps, which is exhibited by the red circles in Figure 2 (c) and (d).

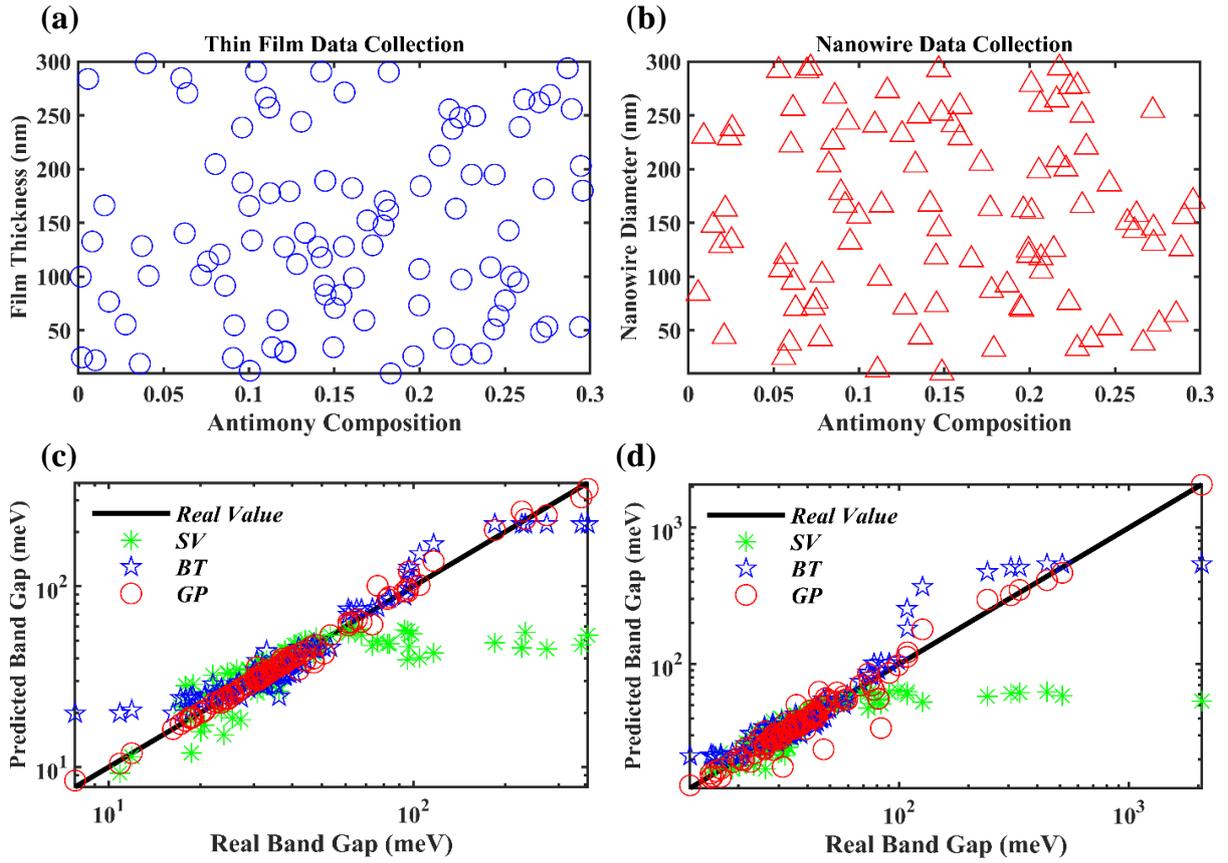

**FIGURE 2:** In the simple task of predicting the *L*-point direct bandgap for trigonal axis oriented **(a)** thin films and **(b)** nanowires, the data points of the training pool are collected from previous works [39-42]. The performances of machine learners using the support vectors (SV), the bagged trees (BT), and the Gaussian processes (GP) are compared in **(c)** and **(d)**, correspondingly.

We then studied a problem with higher level of complexity, where the growth orientation of the Bi$_{1-x}$Sb$_x$ thin film and nanowires is not restricted to the *R3*-symmetry (trigonal) axis, but allowed to be along a general low-symmetry crystalline direction. The ~100 sample data are also collected from the previous works [39-42], which are exhibited in Figure 3 (a) and (b). The distance from the origin to the data point stands for the film thickness in Figure 3 (a) and the nanowire length in Figure 3 (b), while the coordinates in the three-dimensional space formed by the bisectrix, the binary and the trigonal axes, is representing the crystalline orientation along which the thin film or the nanowire is grown. The alloying composition *x* is represented by the size of each data point. As we can see, the sample data are also randomly distributed in the high-dimensional feature space to ensure an unbiased learning and prediction. In Figure 3 (c) and (d), we can see that the general tendency of improvement in the order of SV, BT and GP learners is similar to the cases in Figure 2. However, it is interesting to find out that the overall performance is better than those in Figure 2 (c) and (d). This may be attributed to the preference of higher dimension in the feature space that are commonly observed in machine learning tasks.

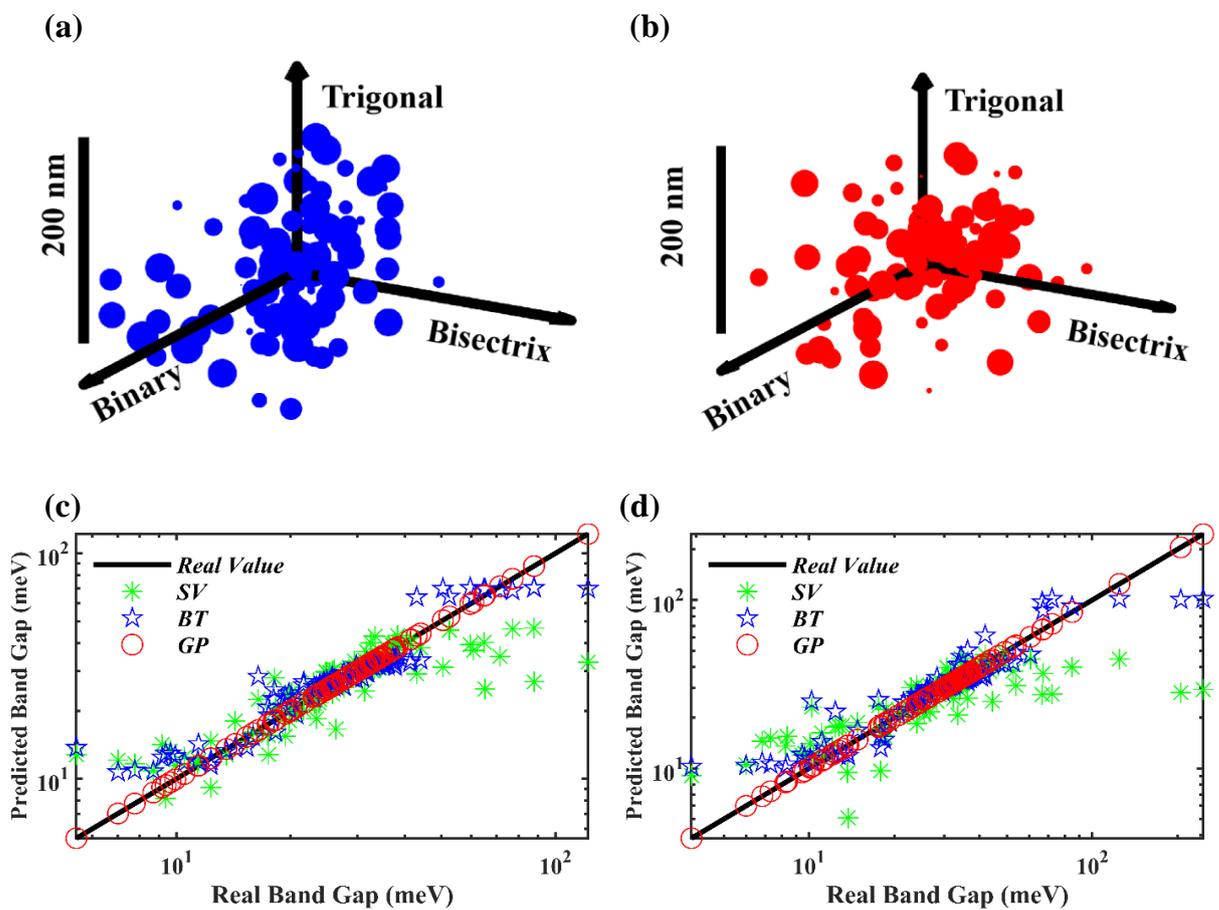

**FIGURE 3:** In our study of (a) thin films and (b) nanowires of arbitrary growth orientations, the data points of the training pool are collected from the previous works [39-42]. The magnitude and the direction of the vector connecting the origin and a specific data point represents the **(a)** film thickness or the **(b)** nanowire diameter, and the **(a)** film or **(b)** wire growth orientation, respectively. The size of the dots stands for the alloying composition $x$. The performances of machine learners using the support vectors (SV), the bagged trees (BT), and the Gaussian processes (GP) are compared in **(c)** and **(d)**, correspondingly.

Since the performance improving tendency of the three algorisms applied in Figure 2 and 3 both implies a potential advantage of machine learners embedded with higher nonlinearity, we now construct a two-layer forward ANN for a further optimization. To investigate how the model complexity affects the prediction, we compared the results obtained by the ANNs with 3, 5, 10 and 20 neurons in the hidden layer for information processing. Their performances are compared in Figure 4 (a) and (b) for the thin film and the nanowire predictors, respectively. It is apparent that the overall prediction accuracy is enhanced. The goodness-of-fit for both the thin films and the nanowires can reach as high as 0.99 when ~10 neurons are used in the hidden layer. An ANN with fewer neurons may suffer from an insufficient information processing, as suggested by the lower goodness-of-fit. However, an ANN with more neurons (e.g. 20), also results in a lower goodness-of-fit, most likely due to the overfitting.

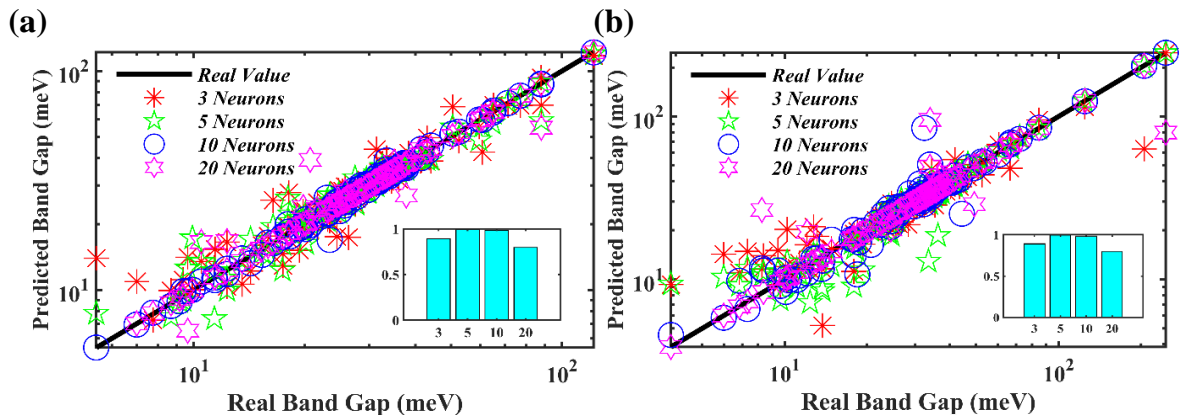

**FIGURE 4:** The prediction performances of the two-layer forward neural network learners for $Bi_{1-x}Sb_x$ **(a)** thin films and **(b)** nanowires, when 3 (red stars), 5 (green pentagrams), 10 (purple circles), and 20 (magenta hexagrams) neurons are used in the hidden layer. The corresponding goodness-of-fit for each machine learner is exhibited in the bar graphs, the decrease of which when too many neurons are used implies an overfitting as we expect

## 4. Conclusions

We have employed the modern machine learning methods to efficiently predict the $L$-point bandgap of the $Bi_{1-x}Sb_x$ thin films and nanowires. This gap is important for the $Bi_{1-x}Sb_x$-based infrared, terahertz, and cryogenic applications, which has been, however, challenging using traditional physics modeling, due to the strong correlation and anomalous quantum confinement. The traditional learners of support vector regression, bagged tree, and Gaussian process regression are used and compared, which implies that stronger nonlinearity should be embedded into the modeling. Then a two-layer forward neural network is built to achieve a goodness-of-fit of as high as ~0.99. The overfitting issue is addressed and avoided by varying the number of neurons in the hidden layer. Beyond the $Bi_{1-x}Sb_x$ system, these machine learning methods can also be used in other alloyed nanomaterials with low crystalline symmetry and strong electronic correlations.

**Acknowledge**

The author(s) acknowledge the Center for Computational Innovations at the Rensselaer Polytechnic Institute for providing the AIMOS supercomputer to support our research and student training.

**Author Contributions**

S.T. designed and performed the research, analyzed the data, and wrote the paper. JJB, SV, AL and AB helped in preparing and organizing the data analysis.

## Conflict of interest

There are no conflicts to declare.

**Table of Contents Entry:**

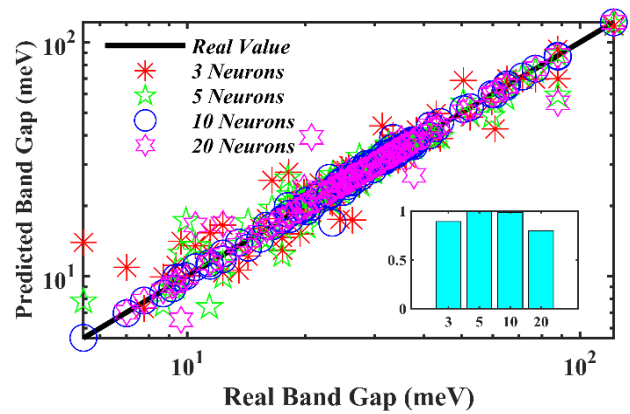

**20-word summary:**

Both traditional and ANN machine learning methods are used to predict $L$-point direct bandgap of $Bi_{1-x}Sb_x$ nanomaterials.